# Survey-Based Analysis of the Proposed Component-Based Development Process

M. Rizwan Jameel Qureshi
Dept. of Computer Science, COMSATS Institute of Information Technology, Lahore
anriz@hotmail.com
Ph # (92-42-5431602) Cell # (03334492203)
M. E. Sandhu
National College of Business Administration & Economics
40 E/I Gulberg III Lahore, Pakistan

## Abstract

*The concept of component-based development (CBD) is widely practiced in software (SW) development. CBD is based on reuse of the existing components with the new ones. The objective of this paper is to propose a novel process model for CBD. Importance of repository has also been discussed. A survey has been conducted to evaluate the proposed model. The results of the survey show that proposed process model can be efficiently implemented for CBD projects.*

## 1. Introduction

Majority of the latest tools help to develop information systems by coping the needs of process models using structured or object oriented (OO) approaches [1,2]. There are lots of attempts by researchers in recent years to adapt and improve component-based development to meet needs of software industry [3,4,5].

The concept of reuse can be explained by taking the example of manufacturing industry. Vehicles manufacturers are so successful because they have used standardized parts. Standardized software components are provided in the form of libraries available with software such as Microsoft Foundation Classes (MFC) and Standard Template Library (STL).

The OO model is the only model among existing process models which provide an artifact explicitly for component based software development (CBD) [6]. These are the attempts made in last few years to propose CBD model [6,7,8,9,10,11,12].

The aim of this paper is to propose a new CBD process model. This is accomplished by using a repository at analysis phase to develop complex systems [13,14]. The proposed model is validated by conducting a survey from SW companies which are dealing globally.

Section 2 proposes a new process model for CBD. Section 3 discusses the function and significance of repository in CBD. Section 4 describes silent features of the new process model. Section 5 describes validation of the proposed model using survey.

## 2. The Proposed Process Model of CBD

Main phases of CBD process model are shown, in figure 1.

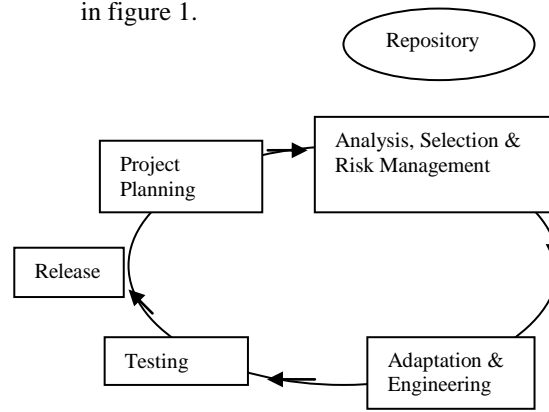

**Figure 1-The Proposed CBD Process Model**

- Project Planning
- Analysis, component selection and Risk Management
- Adaptation & Engineering
- Testing

An initial communication is made with customer to gather main user stories at the start of the project. Project specification is made to prepare cost benefit analysis sheet during the *planning* phase. CBA sheet facilitates to decide that whether SW project is feasible for the customer or not.

*The analysis* phase is started if customer approves the proposal. Detailed user stories are gathered during the analysis phase. A domain analysis helps to accomplish a suitable architecture for the application to be developed. An architectural model of application enables a software engineer to evaluate efficiency of design, judge options of design, and minimize potential threats coupled with software development.

An analyst tries to identify and select those components that can be reused from the components repository during the 'Analysis, Selection & Risk Management' phase. Risks estimations are made about new and existing components. Risk mitigation, monitoring and management (RMMM) plan is made to identify, monitor and manage risks. New components are engineered for those requirements which can not be fulfilled from already developed components.

Reusable components need identification, matching, customization and aggregation. Component matching is made to ensure that the selected component will perform the required functionality, assemble easily into the architecture of new application and possess the quality attributes (e.g., reliability, performance, usability).

The attributes, characteristics and composition among components are identified. Main objective of this phase is to reuse maximum components instead of reinventing the wheel. Reuse also improves output and effectiveness of software engineers.

The reusable components are adapted to meet the user stories of customer and new components are designed and developed during *adaptation & engineering* phase. Component wrapping technique is used to adapt reusable components if programmer is using black box components [6]. The adapted components are integrated and tested during testing phase. The new components are also tested on unit, integration, system and acceptance basis. The SW is evaluated by the customer during the beta testing. The SW is only deployed if customer approves the beta version.

## 3. Significance of Repository in CBD

The purpose of domain engineering is to identify, construct and catalog software components to be reused. The main objective is to establish a set of procedures. These procedures are used to disseminate information among programmers about reusable components using a repository [6,14]. As mentioned in section 2, selection of reusable components is important to improve productivity of component-based software. The repository is used to store and manage reusable components. Main benefits achieved, while working on the reusable components having a repository, are as follows.

- Classification
- Searching
- Modification
- Testing
- Implementation
- Version control
- Change control
- Up to date and consistent documentation

There could be one or more than one repository to select and retrieve components [14]. Repository plays an important role during 'Analysis, Component Selection and Risk Management' phase. Requirements are gathered like traditional software development. The project manager conducted a meeting with all team members. The meeting is called an impact analysis or gap analysis. The meeting is conducted to identify reusable components using a repository. Repository usage at analysis phase helped software engineering team to select strategy that can be used to complete the project. There are three popular strategies adopted for CBD [9,15].

- Components are reused using repository which is populated with all the required components to be reused for the current project.
- Commercially off the shelf components are available to meet the requirements of current project.
- Traditional software development procedures to develop new components

to meet the requirements of current project.

Repository also contained the risk management plan to cater all possible risks which could occur causing the project to fail it. Therefore we can say that risk management is not possible without documentation. Repository is also very helpful to enhance and reengineer the SW in future. This is because documentation eliminates the gap of development. For example, a software house 'ABC' developed a project for a client. The client demanded the documentation (user and system) of the SW. The client company can ask to any SW house such as, 'XYZ' to enhance the SW if documentation is available.

# 4. Main Features of the Proposed CBD Model

These are main features of the proposed model for CBD.

- A new process model in software engineering field.
- Use of repository at analysis phase is a potential benefit.
- Risk management at analysis phase in the new framework to cater the potential risks regarding failure of project.
- The new process model provides strong support for:
    - Reusability
    - Interoperability
    - Upgradeability
    - Less complexity
    - Time saving
    - Cost saving
    - Reliability
    - Improved Quality

# 5. Validation of the Proposed CBD Process Model Using Survey

A survey involving seven software development organizations was conducted to evaluate the proposed CBD process model. A list of software houses was taken from Pakistan Software Export Board [16]. These software houses are developing SW for China, India, Australia, UK and USA. Questionnaire technique was used to gather the data. Thirty eight professionals were selected to fill the questionnaire forms. The people who filled the forms had more than six years experience in software development. Likert scale was ranging from 1 to 5 to gather the data against the questionnaires as shown in Table 1.

| Very low effect | 1 |
|---|---|
| Low effect | 2 |
| Nominal/Average effect | 3 |
| High effect | 4 |
| Very high effect | 5 |

**Table 1-The Range of Likert scale used in Questionnaires**

## 4.1 Data Gathering Technique

A questionnaire was used to evaluate the proposed CBD model. Questionnaire was divided into two main sections. Each section was consisted of different questions. The sections were:

**Questionnaire 1:**
- Suitability of the system development life cycle phases of proposed model for CBD projects.
- Measure the effect of repository for CBD projects.

**Suitability of the System Development Life Cycle Phases of the Proposed Model for CBD Projects**

Table 2 has been created on the basis of evaluations. The parameters evaluated in Table 2 were as follows.

- A- Rank 'Project Planning' phase suitability for component based development (CBD) software projects.
- B- Rank 'Analysis, Selection & Risk Management' phase suitability for CBD projects.
- C- Rank 'Adaptation & Engineering' phase suitability for CBD projects.
- D- Rank 'Testing' phase suitability for CBD projects.

E- Does 'Risk Management' phase cater the potential risks regarding failure of the project?

| Weight → | % of 1 | % of 2 | % of 3 | % of 4 | % of 5 |
|---|---|---|---|---|---|
| Parameter ↓ | | | | | |
| A | | | 26.3 | 55.3 | 18.4 |
| B | | | 13.2 | 63.2 | 23.7 |
| C | | 5.3 | 26.3 | 50 | 18.4 |
| D | | 2.6 | 15.8 | 34.2 | 47.4 |
| E | | 2.6 | 23.7 | 52.6 | 21.1 |

**Table 2- The Suitability of the Proposed**

Table 2 shows that respondents highly supported the proposed model for CBD projects.

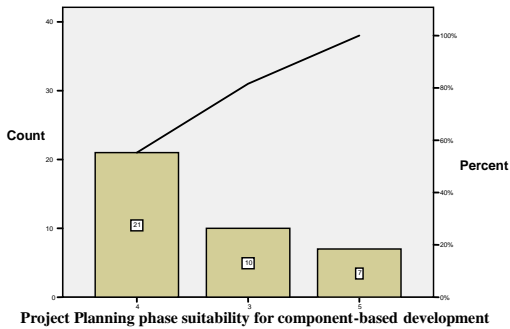

**Figure 2-'Project Planning' Phase Suitability for Component-Based Development (CBD) Software Projects**

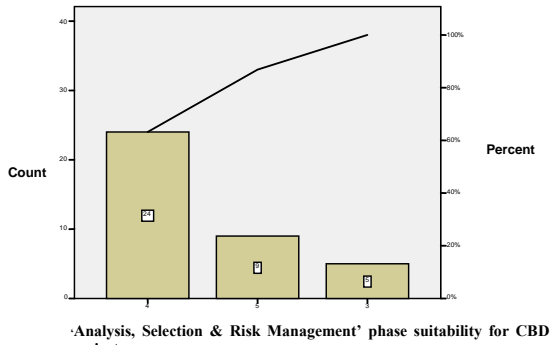

**Figure 3-'Analysis, Selection & Risk Management' Phase Suitability for CBD Projects**

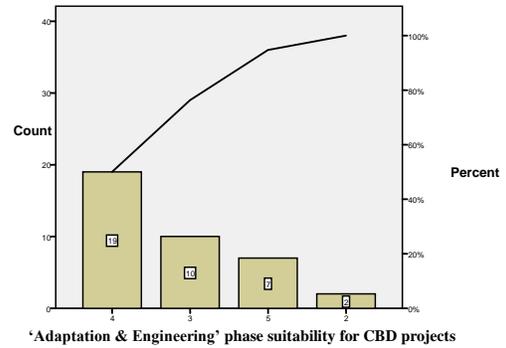

**Figure 4- 'Adaptation & Engineering' Phase Suitability for CBD Projects**

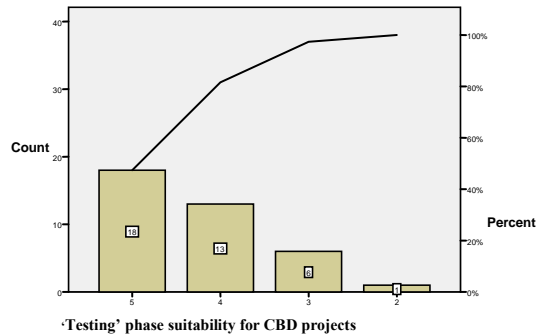

**Figure 5- 'Testing' Phase Suitability for CBD Projects**

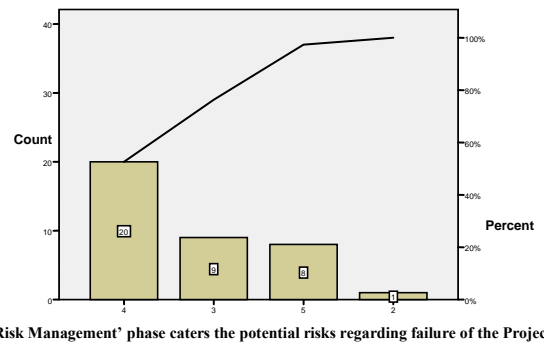

**Figure 6-'Risk Management' Phase Caters the Potential Risks Regarding Failure of the Project**

Figure 2 to 6 shows that the respondents are of the view highly supporting the importance of planning, 'analysis, selection & risk management', 'adaptation & engineering' and testing phases for CBD projects. These results indicate that the proposed changes in the SDLC of CBD projects are validating the implementation of the case studies conducted in one of the seven software organizations. These results also show that the proposed CBD model is better than convential OO process models.

**Effect of Repository on the Proposed CBD Process Model**

Table 3 has been created on the basis of evaluations. The parameters evaluated in Table 3 were as follows.

- A- Repository needs at Analysis Selection & Risk Management Phase.
- B- Repository helps to classify reusable components.
- C- Repository makes easier to search reusable components.
- D- Repository facilitates modification of reusable components.
- E- Repository helps to test reusable components.
- F- Repository facilitates implementation of reusable components.
- G- Repository makes it easier to manage versions of reusable components.
- H- Repository helps to maintain up to date and consistent documentation.

| Weight → Parameters ↓ | % of 1 | % of 2 | % of 3 | % of 4 | % of 5 |
|---|---|---|---|---|---|
| A | 2.6 | 7.9 | 13.2 | 47.4 | 28.9 |
| B |  | 5.3 | 18.4 | 50 | 26.3 |
| C |  | 5.3 | 21.1 | 39.5 | 34.2 |
| D |  | 13.2 | 15.8 | 52.6 | 18.4 |
| E |  | 10.5 | 26.3 | 55.3 | 7.9 |
| F |  |  | 36.8 | 42.1 | 21.1 |
| G |  | 10.5 | 21.1 | 26.3 | 42.1 |
| H |  | 5.3 | 15.8 | 36.8 | 42.1 |

**Table 3-Effect of Repository on the Proposed CBD Process Model**

The results of Table 3 show that repository has high effect on the proposed process model for CBD projects.

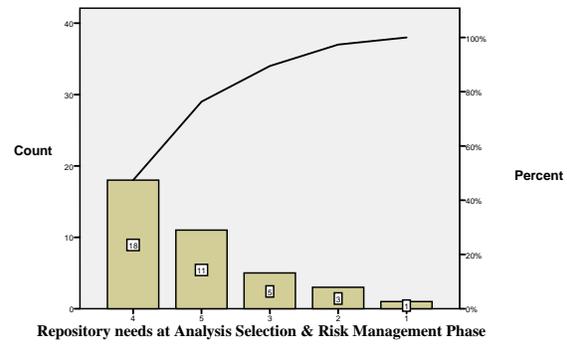

**Figure 7-Repository needs at Analysis Selection & Risk Management Phase**

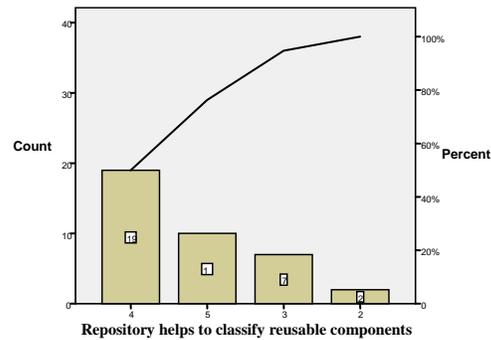

**Figure 8-Repository Helps to Classify Reusable Components**

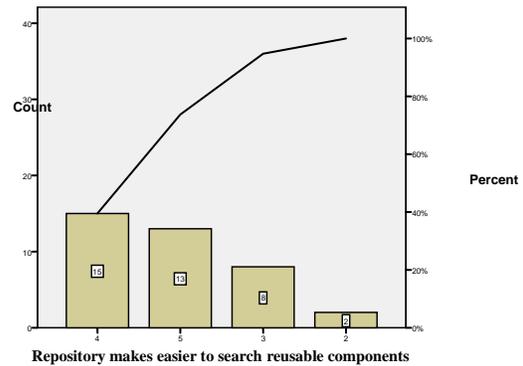

**Figure 9- Repository Makes Easier to Search Reusable Components**

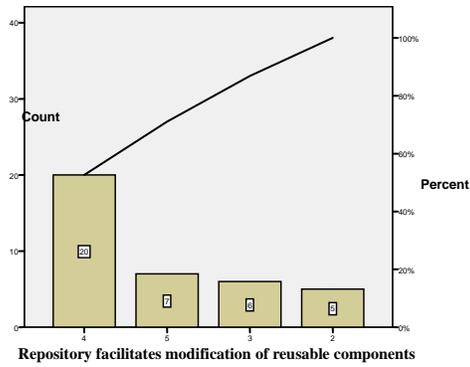

**Figure 10-Repository Facilitates Modification of Reusable Components**

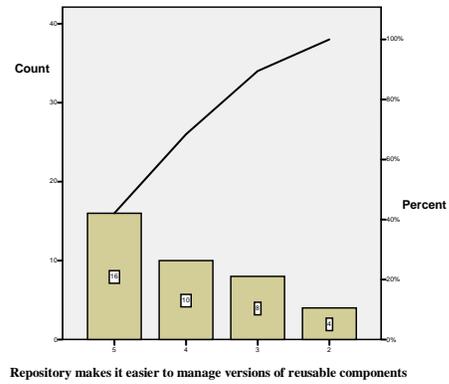

**Figure 13 Repository Makes it Easier to Manage Versions of Reusable Components**

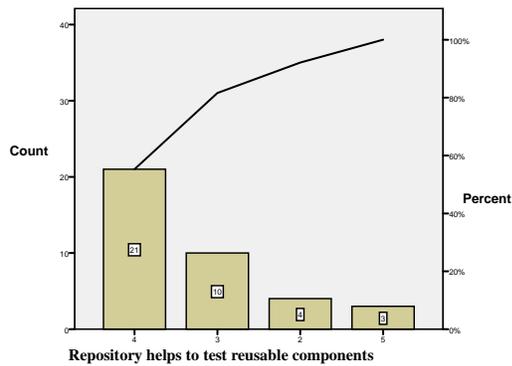

**Figure 11-Repository Helps To Test Reusable Components**

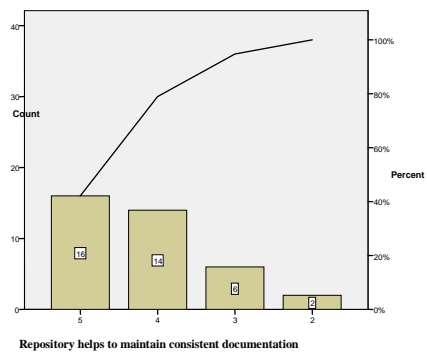

**Figure 14-Repository Helps to Maintain Up to Date and Consistent Documentation**

Figure 7 to 14 shows effect of repository on the proposed process model. It shows that most of SW developers are of the view that repository has high effect on the development of CBD projects.

## 5. Conclusion

This paper supports practice of CBD instead of traditional software development. A process model has been presented for the Component Based SW Engineering (CBSE). Role and importance of repository in CBD has also been discussed. The proposed model is validated by conducting a survey from seven software companies which are dealing globally. From this validation it is concluded that the proposed CBD model is highly supported by majority of the SW developers.

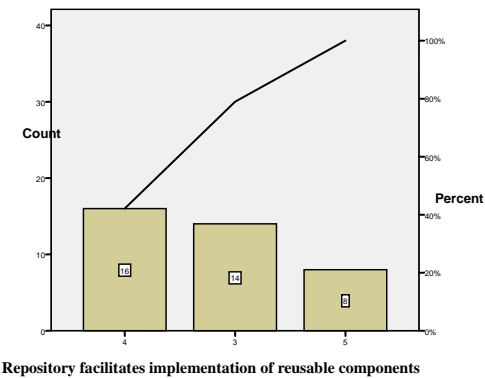

**Figure 12- Repository Facilitates Implementation of Reusable Components**